\documentstyle[aps,graphics,epsfig]{revtex}
\newcommand{\bfm}[1]{\mbox{\boldmath${#1}$}}
\begin{document}
\draft

\title{A new one parameter
deformation of the exponential function}
\author{G. Kaniadakis, and A.M. Scarfone\footnote{electronic address: kaniadakis@polito.it, scarfone@polito.it
}}
\address{Dipartimento di Fisica - Politecnico di Torino -
Corso Duca degli Abruzzi 24, 10129 Torino, Italy \\
Istituto Nazionale di Fisica della Materia - Unit\'a del
 Politecnico di Torino }

\date{\today}
\maketitle
\begin {abstract}

Recently, in the ref. Physica A \bfm{296} 405 (2001), a new one
parameter deformation for the exponential function
$\exp_{_{\{{\scriptstyle \kappa}\}}}\!(x)=
\left(\sqrt{1+\kappa^2x^2}+\kappa\,x\right)^{1/\kappa}; \
\exp_{_{\{{\scriptstyle 0}\}}}\!(x)=\exp\,(x)$, which presents a
power law asymptotic behaviour, has been proposed. The statistical
distribution $f=Z^{-1}\exp_{_{\{{\scriptstyle
\kappa}\}}}[-\beta(E-\mu)]$, has been obtained both as stable
stationary state of a proper non linear kinetics and as the state
which maximizes a new entropic form. In the present contribution,
starting from the $\kappa$-algebra and after introducing the
$\kappa$-analysis, we obtain the $\kappa$-exponential
$\exp_{_{\{{\scriptstyle \kappa}\}}}(x)$ as the eigenstate of the
$\kappa$-derivative and study its main mathematical properties.
\end {abstract}

\section{Introduction}

The Maxwell-Boltzman (MB) statistical distribution
$f=Z^{-1}\exp[-\beta(E-\mu)]$ is constructed from the exponential
function which imposes its main features. There are several
experimental evidences in various physical systems of distribution
functions with a behaviour different from the one of the MB
distribution. Most of the observed deviations are related with the
tails of the distribution which decay slower than an exponential,
following a power law \cite{Tribelsky,Lih,Naudts}.

A way to obtain new distribution functions consists in modifying
properly the exponential function \cite{Tsallis,Borges}. Recently,
in the refs. \cite{Kaniadakis1,Kaniadakis2}, the new, one
parameter deformation for the exponential function has been
proposed:
\begin{equation}
\exp_{_{\{{\scriptstyle \kappa}\}}}(x)=
\left(\sqrt{1+\kappa^2x^2}+\kappa\,x\right)^{1/\kappa} \
.\label{ek}
\end{equation}

We recall briefly the main properties of the $\kappa$-exponential
$\exp_{_{\{{\scriptstyle \kappa}\}}}(x)=\exp_{_{\{{\scriptstyle
-\kappa}\}}}(x)$ which reduces to the standard one as
$\kappa\rightarrow 0$, namely $\exp_{_{\{{0}\}}}(x)=\exp x$, and
shows a power law asymptotic behaviour
\begin{equation}
\exp_{_{\{{\scriptstyle
\kappa}\}}}(x){\atop\stackrel{\textstyle\sim}{\scriptstyle
x\rightarrow \pm \infty}}|2\kappa x|^{\pm 1/|\kappa|} \ .
\end{equation}
The $\kappa$-exponential
decreases for $x\rightarrow -\infty$ and increases for
$x\rightarrow +\infty$ with the same rapidness
\begin{equation}\exp_{_{\{{\scriptstyle \kappa}\}}}(x)
\exp_{_{\{{\scriptstyle \kappa}\}}}(-x)=1 \ ,\label{rp}
\end{equation}
is a positive $\exp_{_{\{{\scriptstyle \kappa}\}}}(x)\in {\bfm
R}^+$, and increasing  $d \exp_{_{\{{\scriptstyle
\kappa}\}}}(x)/dx>0$ function $\forall x, \kappa \in {\bfm R}$ and
obeys to the scaling law
\begin{equation}
[\exp_{_{\{{\scriptstyle
\kappa}\}}}(x)]^{\,\lambda}=\exp_{_{\{{\scriptstyle
\kappa/\lambda}\}}}(\lambda x) \ .
\end{equation}

The inverse function, namely the $\kappa$-logarithm, is defined
through
\begin{equation}
\ln_{_{\{{\scriptstyle \kappa}\}}}(x)=
\frac{x^{\kappa}-x^{-\kappa}}{2\,\kappa} \ .\label{N28}
\end{equation}
The $\kappa$-logarithm is a real and increasing function $\forall
x\in{\bfm R}^+$. We have that $ \ln_{_{\{{0}\}}}(x)=\ln x$ and
$\ln_{_{\{{\scriptstyle \kappa}\}}}(x)=\ln_{_{\{{\scriptstyle
-\kappa}\}}}(x)$. Its asymptotic behaviour is given by
\begin{eqnarray}
\ln_{_{\{{\scriptstyle
\kappa}\}}}(x){\atop\stackrel{\textstyle\sim}{\scriptstyle
x\rightarrow {\,0^+}}}-\frac{1}{|2\kappa|}\,\,x^{-|\kappa|} \ \ \
\ \ ; \ \ \ \ \ \ln_{_{\{{\scriptstyle
\kappa}\}}}(x){\atop\stackrel{\textstyle\sim}{\scriptstyle
x\rightarrow +\infty}}\frac{1}{|2\kappa|}\,\,x^{|\kappa|} \ .
\end{eqnarray}
Finally, the following scaling law holds
\begin{equation}
\ln_{_{\{{\scriptstyle
\kappa}\}}}(x^{\lambda})=\lambda\,\ln_{_{\{{\scriptstyle \lambda
\kappa}\}}}(x) \ ,
\end{equation}
from which for $\lambda=-1$ we obtain:
\begin{eqnarray}
\ln_{_{\{{\scriptstyle \kappa}\}}}(1/f)=-\ln_{_{\{{\scriptstyle
\kappa}\}}}(f) \ .\label{lk}
\end{eqnarray}

In Figure 1, the plot of the $\kappa$-exponential given by Eq.
(\ref{ek}) for different values of the parameter $\kappa$ is
reported. The curve with $\kappa=0$ corresponds to the standard
exponential.

The distribution function
\begin{equation}
f=\frac{1}{Z}\exp_{_{\{{\scriptstyle \kappa}\}}}[-\beta(E-\mu)] \
, \label{dis}
\end{equation}
which reduces to the MB one as $\kappa\rightarrow0$, has been
obtained \cite{Kaniadakis1} both as stable stationary state of a
non linear kinetics and as the state which maximizes the new
entropic form properly constrained
\begin{equation}
S_{\kappa}=\int d^n v
\,[\,c(\kappa)f^{1+\kappa}+c(-\kappa)f^{1-\kappa}] \ \ \ \ ; \ \ \
\ c(\kappa)=-\frac{Z^{\kappa}}{2\,\kappa\,(1+\kappa)} \ .
\end{equation}

In Figure 2, the plot of the distribution function given by Eq.
(\ref{dis}), for Brownian particles where $E=m\,v^2/2$ and
$\mu=0$, vs the dimensionless velocity is reported. The curve with
$\kappa=0$ corresponds to the standard MB distribution.

\section{Algebraic Structures}
Let us consider the sum $\stackrel{\kappa}{\oplus}$ between
elements of the set $X\equiv{\bfm R}$ \ introduced in ref.
\cite{Kaniadakis1}
\begin{equation}
x\stackrel{\kappa}{\oplus}z=x\sqrt{1+\kappa^2z^2}+z\sqrt{1+\kappa^2x^2}
\ ,\label{ksum}
\end{equation}
and rewrite Eq.(\ref{ksum}) under the following equivalent form
\begin{eqnarray}
x\stackrel{\kappa}{\oplus}z={1\over\kappa}\,\sinh\left(\,{\rm
arcsinh}\,\kappa x+{\rm arcsinh}\,\kappa z\,\right) \
.\label{ksum1}
\end{eqnarray}

Of course we have that $x \stackrel{0}{\oplus}z= x+z$ and all the
properties of the standard undeformed sum still hold:\\ i)
associative law: $(x \stackrel{\kappa}{\oplus}z)
 \stackrel{\kappa}{\oplus}t=x
 \stackrel{\kappa}{\oplus}(z
 \stackrel{\kappa}{\oplus}t)$;\\
ii) neutral element: $x \stackrel{\kappa}{\oplus}0= 0
 \stackrel{\kappa}{\oplus}x= x$;\\
iii) opposite element: $x
\stackrel{\kappa}{\oplus}(-x)=(-x)\stackrel{\kappa}{\oplus}x=
0$;\\ iv) commutative law: $x
 \stackrel{\kappa}{\oplus}z=z
 \stackrel{\kappa}{\oplus}x$.\\
Thus the algebraic structure $(X,\, \stackrel{\kappa}{\oplus})$
forms an abelian group.\\ The difference
$\stackrel{\kappa}{\ominus}$ can be defined through $x
\stackrel{\kappa}{\ominus}z= x \stackrel{\kappa}{\oplus}(-z)$ and
the identity \hbox{$(x\stackrel{\kappa}{\oplus}z)\,(
x\stackrel{\kappa}{\ominus}z)=x^2-z^2$} holds.

Let us consider the following product $\stackrel{\kappa}{\otimes}$
between elements of the set $X$
\begin{eqnarray}
x\stackrel{\kappa}{\otimes}z={1\over\kappa}\,\sinh\left[\,{1\over\kappa}\,{\rm
arcsinh}\,\kappa x\cdot{\rm arcsinh}\,\kappa z\,\right] \
,\label{kprod}
\end{eqnarray}
which reduces to the standard product when the parameter $\kappa$
approaches to zero, namely $x \stackrel{0}{\otimes}z= x\cdot z$.

The product $\stackrel{\kappa}{\otimes}$ has the following
properties:\\ i) associative law: $(x \stackrel{\kappa}{\otimes}z)
 \stackrel{\kappa}{\otimes}t=x
 \stackrel{\kappa}{\otimes}(z
 \stackrel{\kappa}{\otimes}t)$;\\
 ii) neutral element: $x \stackrel{\kappa}{\otimes}I=I
 \stackrel{\kappa}{\otimes}x= x$;\\
iii) inverse element:
  $x \stackrel{\kappa}{\otimes}\overline x=
\overline x \stackrel{\kappa}{\otimes}x=I$;\\ iv) commutative law:
$x \stackrel{\kappa}{\otimes}z= z
 \stackrel{\kappa}{\otimes}x$,\\
being  $I=\kappa^{-1}\,\sinh\kappa$ and $\overline
x=\kappa^{-1}\,\sinh(\kappa^2/{\rm arcsinh}\,\kappa x)$. Thus the
algebraic structure $(X,\, \stackrel{\kappa}{\otimes})$ forms an
abelian group. The division $\stackrel{\kappa}{\oslash}$ can be
defined as $x \stackrel{\kappa}{\oslash}z=x
\stackrel{\kappa}{\otimes}\overline z$.

Observe that the sum $\stackrel{\kappa}{\oplus}$ and product
$\stackrel{\kappa}{\otimes}$ obey the distributive law $
t\stackrel{\kappa}{\otimes}(x \stackrel{\kappa}{\oplus}z)=(t
\stackrel{\kappa}{\otimes}x) \stackrel{\kappa}{\oplus}(t
\stackrel{\kappa}{\otimes}z)$ and then we can conclude that the
algebraic structure
$(X,\,\stackrel{\kappa}{\oplus},\,\stackrel{\kappa}{\otimes})$ is
a field. When the deformation parameter $\kappa$ approaches to
zero we reobtain the well known undeformed algebraic structure
$(X,\, + ,\, \cdot\,)$ which forms the standard field of the real
numbers, being $+$ and $\cdot$ the standard sum and product,
respectively. Finally, for the sum $\stackrel{\kappa}{\oplus}$ and
the ordinary product $\cdot$ holds the pseudodistributive law
$t\cdot(x \stackrel{\kappa}{\oplus} z)=(t\cdot x)
\stackrel{\kappa/t}{\oplus} (t\cdot z)$ so that the structure
$(X,\,\stackrel{\kappa}{\oplus},\,\cdot\,)$ forms a pseudofield.
\vspace{2mm}

We remark that, after introducing:
\begin{eqnarray}
x_{_{\{{\scriptstyle \kappa}\}}}=\frac{1}{\kappa}\, {\rm arcsinh}
\,\kappa\,x \ ,\label{xk}
\end{eqnarray}
Eqs. (\ref{ksum1}) and (\ref{kprod}) can be written as
\begin{equation}
x_{_{\{{\scriptstyle \kappa}\}}}+z_{_{\{{\scriptstyle
\kappa}\}}}=(x \stackrel{\kappa}{\oplus}z )_{_{\{{\scriptstyle
\kappa}\}}} \ ,
\end{equation}
\begin{equation}
x_{_{\{{\scriptstyle \kappa}\}}}\cdot z_{_{\{{\scriptstyle
\kappa}\}}}=(x \stackrel{\kappa}{\otimes}z )_{_{\{{\scriptstyle
\kappa}\}}} \ ,
\end{equation}
and then we can conclude that an isomorphism between $(X,\,
\stackrel{\kappa}{\oplus},\, \stackrel{\kappa}{\otimes})$ and
$(X,\, + ,\, \cdot\,)$ exists.
\section{The \boldmath$\kappa$-derivative}

Consider the two algebric structure
$(X,\,\stackrel{\kappa}{\oplus},\,\cdot\,)$ and $(Y,\, +
,\,\cdot\,)$ with $X=\{x,z,t,...\}\equiv{\bfm R}$ and
$Y=\{y,u,v,...\}\equiv{\bfm R}$. Let us introduce the set of the
direct functions ${\cal F}=\{f: X\stackrel{f}\rightarrow Y \}$
with ${\cal F}\subseteq C^{\infty}(X)$ and the set of the inverse
functions ${\cal G}=\{g: Y\stackrel{g}\rightarrow X \}$ with
${\cal G}\subseteq C^{\infty}(Y)$.

We define the $\kappa$-derivative for the functions of the set
${\cal F}$ through
\begin{equation}
\frac{d\,f(x)}{d_{_{\{{\scriptstyle \kappa}\}}}
x}=\lim_{z\rightarrow x}\frac{f (x)-f(z)}{\displaystyle{ x
\stackrel{\kappa}{\ominus}z}}\ \ ,
\end{equation}
with $x,z\in X$ and $f(x),f(z)\in Y$. The $\kappa$-differential $d
_{_{\{{\scriptstyle \kappa}\}}}x$ is given by
\begin{eqnarray}
d _{_{\{{\scriptstyle \kappa}\}}}x=\lim_{z\rightarrow x} x
\stackrel{\kappa}{\ominus}z= \lim_{z\rightarrow x}
\frac{x^2-z^2}{\displaystyle{x\stackrel{\kappa}{\oplus}z}}=
\frac{d\,x}{\displaystyle{\sqrt{1+\kappa^2\,x^2} }} \ .
\end{eqnarray}
It is easy to verify that $d _{_{\{{\scriptstyle
\kappa}\}}}x=dx_{_{\{{\scriptstyle \kappa}\}}}$, being
$x_{_{\{{\scriptstyle \kappa}\}}}$ given by Eq. (\ref{xk}). We
observe that the ${\kappa}$-derivative, which reduces to the usual
one as the deformation parameter ${\kappa}\rightarrow 0$, can be
written in the form
\begin{equation}
\frac{d \, f(x)}{d _{_{\{{\scriptstyle \kappa}\}}}x}=\frac{d \,
f(x)}{d \, x_{_{\{{\scriptstyle
\kappa}\}}}}=\sqrt{1+\kappa^2\,x^2}\,\,\frac{d \, f(x)}{d \, x} \
,
\end{equation}
from which it appears clearly that the ${\kappa}$-calculus is
governed by the same rules of the ordinary one.

The $\kappa$-exponential $\exp_{_{\{{\scriptstyle \kappa}\}}}(x)$
given by Eq. (\ref{ek}) can be defined as eigenstate of the
$\kappa$-derivative for the direct functions
\begin{equation}
\frac{d\,\exp_{_{\{{\scriptstyle \kappa}\}}}(x)}{d\,
x_{_{\{{\scriptstyle \kappa}\}}}}=\exp_{_{\{{\scriptstyle
\kappa}\}}}(x) \ .
\end{equation}

Analogously the $\kappa$-derivative of the functions of the set
${\cal G}$ is defined through
\begin{equation}
\frac{d_{_{\{{\scriptstyle
\kappa}\}}}\,\,g(y)}{d\,y}=\frac{d\,\,\, g(y)_{_{\{{\scriptstyle
\kappa}\}}}}{d \, y}=\lim_{u\rightarrow
y}\frac{g(y)\stackrel{\kappa}{\ominus}g(u)}{\displaystyle{y-u}}=
{1\over\sqrt{1+\kappa^2\,g(y)^2}}\,\frac{d\,g(y)}{d \, y} \ ,
\end{equation}
with $y,u\in Y$ and $g(y),g(u)\in X$.

The $\kappa$-logarithm can be viewed as the inverse function of
the $\kappa$-exponential and can also be defined using the
$\kappa$-derivative for the inverse functions through
\begin{equation}
\frac{d_{_{\{{\scriptstyle \kappa}\}}} \, \ln_{_{\{{\scriptstyle
\kappa}\}}}(y)}{d \, y}=\frac{1}{y} \ .
\end{equation}

We recall that in ref. \cite{Kaniadakis1}, starting from the
$\kappa$-exponential and after defining the $\kappa$-sine and
$\kappa$-cosine, the $\kappa$-hyperbolic and $\kappa$-cyclic
trigonometries have been introduced. The inverse functions can be
introduced also from the $\kappa$-logarithm so that a
$\kappa$-mathematics can be constructed naturally which reduces to
the standard one as the deformation parameter approaches to zero.
\section{Composition of Probabilities }

The previously introduced $\kappa$-sum permits us to write the
following relevant property of the $\kappa$-exponential
\begin{equation}
\exp_{_{\{{\scriptstyle \kappa}\}}}(x) \exp_{_{\{{\scriptstyle
\kappa}\}}}(z) =\exp_{_{\{{\scriptstyle \kappa}\}}}( x
\stackrel{\kappa}{\oplus}z ) \ , \label{N188}
\end{equation}
which, when $z=-x$, reduces to Eq. (\ref{rp}). From Eq.
(\ref{N188}) the following property of $\kappa$-logarithm
immediately results
\begin{eqnarray}
\ln_{_{\{{\scriptstyle \kappa}\}}}(f \cdot
h)=\ln_{_{\{{\scriptstyle \kappa}\}}}(f)\stackrel{\kappa}{\oplus}
\ln_{_{\{{\scriptstyle \kappa}\}}}(h) \ , \label{N23}
\end{eqnarray}
which, in the case $h=1/f$, reproduces Eq. (\ref{lk}).

Commonly $\exp_{_{\{{\scriptstyle \kappa}\}}}(x)$ represents a
probability distribution function while $x$ is a dimensionless
energy. Then, in Eq. (\ref{N188}), the standard composition of
probabilities with a deformed additivity of energies appear. In
the following we show that it is possible to preserve the standard
additivity for the energies if we modify properly the composition
rule of the probabilities.

For simplicity we limit ourself to consider the subset of positive
normalized functions ${\cal D}=\{f,h,w,...\}$. Following ref.
\cite{Kaniadakis2}, we introduce the new product
$\otimes\mbox{\raisebox{-2mm}{\hspace{-2.2mm}$\scriptstyle
\kappa$}} \hspace{1mm}$ through
\begin{eqnarray}
f\otimes\mbox{\raisebox{-2mm}{\hspace{-3mm}$\scriptstyle \kappa$}}
\hspace{2mm}h= \exp_{_{\{{\scriptstyle
\kappa}\}}}\!\left(\,\ln_{_{\{{\scriptstyle
\kappa}\}}}f+\ln_{_{\{{\scriptstyle \kappa}\}}}h\right) \ ,
\label{N04}
\end{eqnarray}
which reduces to the ordinary product as $\kappa\rightarrow 0$,
namely
$f\otimes\mbox{\raisebox{-2.3mm}{\hspace{-2.7mm}$\scriptstyle 0$}}
\hspace{2mm}h= f\cdot h$.

The product
$\otimes\mbox{\raisebox{-2mm}{\hspace{-2.2mm}$\scriptstyle
\kappa$}} \hspace{1mm}$ has the same properties of the ordinary
one:\\ i) associative law:
$(f\otimes\mbox{\raisebox{-2mm}{\hspace{-3mm}$\scriptstyle
\kappa$}}
\hspace{2mm}h)\otimes\mbox{\raisebox{-2mm}{\hspace{-3mm}$\scriptstyle
\kappa$}}
\hspace{2mm}w=f\otimes\mbox{\raisebox{-2mm}{\hspace{-3mm}$\scriptstyle
\kappa$}} \hspace{2mm}
(h\otimes\mbox{\raisebox{-2mm}{\hspace{-3mm}$\scriptstyle
\kappa$}} \hspace{2mm}w)$;\\ ii) neutral element:
$f\otimes\mbox{\raisebox{-2mm}{\hspace{-3mm}$\scriptstyle
\kappa$}}
\hspace{2mm}1=1\otimes\mbox{\raisebox{-2mm}{\hspace{-3mm}$\scriptstyle
\kappa$}} \hspace{2mm}f=f$;\\ iii) inverse element:
$f\otimes\mbox{\raisebox{-2mm}{\hspace{-3mm}$\scriptstyle
\kappa$}} \hspace{2mm}(1/f)=
(1/f)\otimes\mbox{\raisebox{-2mm}{\hspace{-3mm}$\scriptstyle
\kappa$}} \hspace{2mm}f=1$;\\ iv) commutative law: $f
\otimes\mbox{\raisebox{-2mm}{\hspace{-3mm}$\scriptstyle \kappa$}}
\hspace{2mm}h=h\otimes\mbox{\raisebox{-2mm}{\hspace{-3mm}$\scriptstyle
\kappa$}} \hspace{2mm}f$.\\ Thus the algebraic structure $({\cal
D},\,\otimes\mbox{\raisebox{-2mm}{\hspace{-2mm}$\scriptstyle
\kappa$}} \hspace{2mm})$ forms an abelian group. The division
$\oslash\mbox{\raisebox{-2mm}{\hspace{-2.2mm}$\scriptstyle
\kappa$}}\hspace{.5mm}$ can be defined through\\
$f\oslash\mbox{\raisebox{-2mm}{\hspace{-3mm}$\scriptstyle
\kappa$}} \hspace{2mm}h=f
\otimes\mbox{\raisebox{-2mm}{\hspace{-3mm}$\scriptstyle \kappa$}}
\hspace{1mm}(1/h)$. Finally,
$f\otimes\mbox{\raisebox{-2mm}{\hspace{-3mm}$\scriptstyle
\kappa$}}
\hspace{2mm}0=0\otimes\mbox{\raisebox{-2mm}{\hspace{-3mm}$\scriptstyle
\kappa$}} \hspace{2mm}f= 0$.

It is easy to verify that the product
$\otimes\mbox{\raisebox{-2mm}{\hspace{-2.2mm}$\scriptstyle
\kappa$}} \hspace{1mm}$ and the sum
$\oplus\mbox{\raisebox{-2mm}{\hspace{-2mm}$\scriptstyle \kappa$}}
\hspace{1mm}$ defined through :
\begin{eqnarray}
f\oplus\mbox{\raisebox{-2mm}{\hspace{-3mm}$\scriptstyle \kappa$}}
\hspace{2mm}h=\exp_{_{\{{\scriptstyle
\kappa}\}}}\left\{\ln\Big[\exp\left(\ln_{_{\{{\scriptstyle
\kappa}\}}}f\right)+ \exp\left(\ln_{_{\{{\scriptstyle
\kappa}\}}}h\right)\Big]\right\} \ ,
\end{eqnarray}
are distributive
$w\otimes\mbox{\raisebox{-2mm}{\hspace{-3mm}$\scriptstyle
\kappa$}}
\hspace{2mm}(f\oplus\mbox{\raisebox{-2mm}{\hspace{-3mm}$\scriptstyle
\kappa$}} \hspace{2mm}h)=
(w\otimes\mbox{\raisebox{-2mm}{\hspace{-3mm}$\scriptstyle
\kappa$}} \hspace{2mm}f)
\oplus\mbox{\raisebox{-2mm}{\hspace{-3mm}$\scriptstyle \kappa$}}
\hspace{2mm}(w\otimes\mbox{\raisebox{-2mm}{\hspace{-3mm}$\scriptstyle
\kappa$}} \hspace{2mm}h)$.

The sum $\oplus\mbox{\raisebox{-2mm}{\hspace{-2mm}$\scriptstyle
\kappa$}} \hspace{1mm}$, for which results $f
\oplus\mbox{\raisebox{-2.4mm}{\hspace{-3mm}$\scriptstyle0$}}
\hspace{2mm}h=f+h$, has the following properties:\\ i) associative
law: $(f\oplus\mbox{\raisebox{-2mm}{\hspace{-3mm}$\scriptstyle
\kappa$}} \hspace{2mm}h)
\oplus\mbox{\raisebox{-2mm}{\hspace{-3mm}$\scriptstyle \kappa$}}
\hspace{2mm}w=f\oplus\mbox{\raisebox{-2mm}{\hspace{-3mm}$\scriptstyle
\kappa$}} \hspace{2mm}(h
\oplus\mbox{\raisebox{-2mm}{\hspace{-3mm}$\scriptstyle \kappa$}}
\hspace{2mm}w)$;\\ ii) neutral element:
$f\oplus\mbox{\raisebox{-2mm}{\hspace{-3mm}$\scriptstyle \kappa$}}
\hspace{2mm}0=0
\oplus\mbox{\raisebox{-2mm}{\hspace{-3mm}$\scriptstyle \kappa$}}
\hspace{2mm}f=f$;\\ iii) commutative law:
$f\oplus\mbox{\raisebox{-2mm}{\hspace{-3mm}$\scriptstyle \kappa$}}
\hspace{2mm}h=h\oplus\mbox{\raisebox{-2mm}{\hspace{-3mm}$\scriptstyle
\kappa$}} \hspace{2mm}f$.

The product
$\otimes\mbox{\raisebox{-2mm}{\hspace{-2mm}$\scriptstyle \kappa$}}
\hspace{1mm}$ permits us to write the following property of the
$\kappa$-exponential
\begin{eqnarray} \exp_{_{\{{\scriptstyle
\kappa}\}}}(x)\otimes\mbox{\raisebox{-2mm}{\hspace{-3mm}$\scriptstyle
\kappa$}} \hspace{2mm} \!\exp_{_{\{{\scriptstyle
\kappa}\}}}(z)=\exp_{_{\{{\scriptstyle \kappa}\}}}(x+z) \ ,
\label{N19}
\end{eqnarray}
which, for $z=-x$, reduces to Eq. (\ref{rp}). We note that in Eq.
(\ref{N19}) it appears the standard sum of dimensionless energies
together the new deformed product for the probability density
distributions. Eq. (\ref{N19}) can be written also in the form
\begin{eqnarray}
\ln_{_{\{{\scriptstyle
\kappa}\}}}(f\otimes\mbox{\raisebox{-2mm}{\hspace{-3mm}$\scriptstyle
\kappa$}} \hspace{2mm}h)=\ln_{_{\{{\scriptstyle \kappa}\}}}(f)+
\ln_{_{\{{\scriptstyle \kappa}\}}}(h) \ .\label{N24}
\end{eqnarray}
Eq. (\ref{N24}) gives a relevant property for the
$\kappa$-logaritm and in the case $h=1/f$, reduces to Eq.
(\ref{lk}).

\vspace{20mm}

\begin{center}
\bf Figure captions
\end{center}

{\bf Fig.1} Plot of the $\kappa$-exponential given by Eq.
(\ref{ek}) for different values of $\kappa$.

\hspace{9mm} The standard exponential corresponds to $\kappa=0$.\\

{\bf Fig.2} Plot of the distribution function given by Eq.
(\ref{dis}) with $E=m\,v^2/2$ and $ \mu=0$, vs the dimensionless
velocity

\hspace{9mm} for different values of $\kappa$. The standard
exponential corresponds to $\kappa=0$.\\

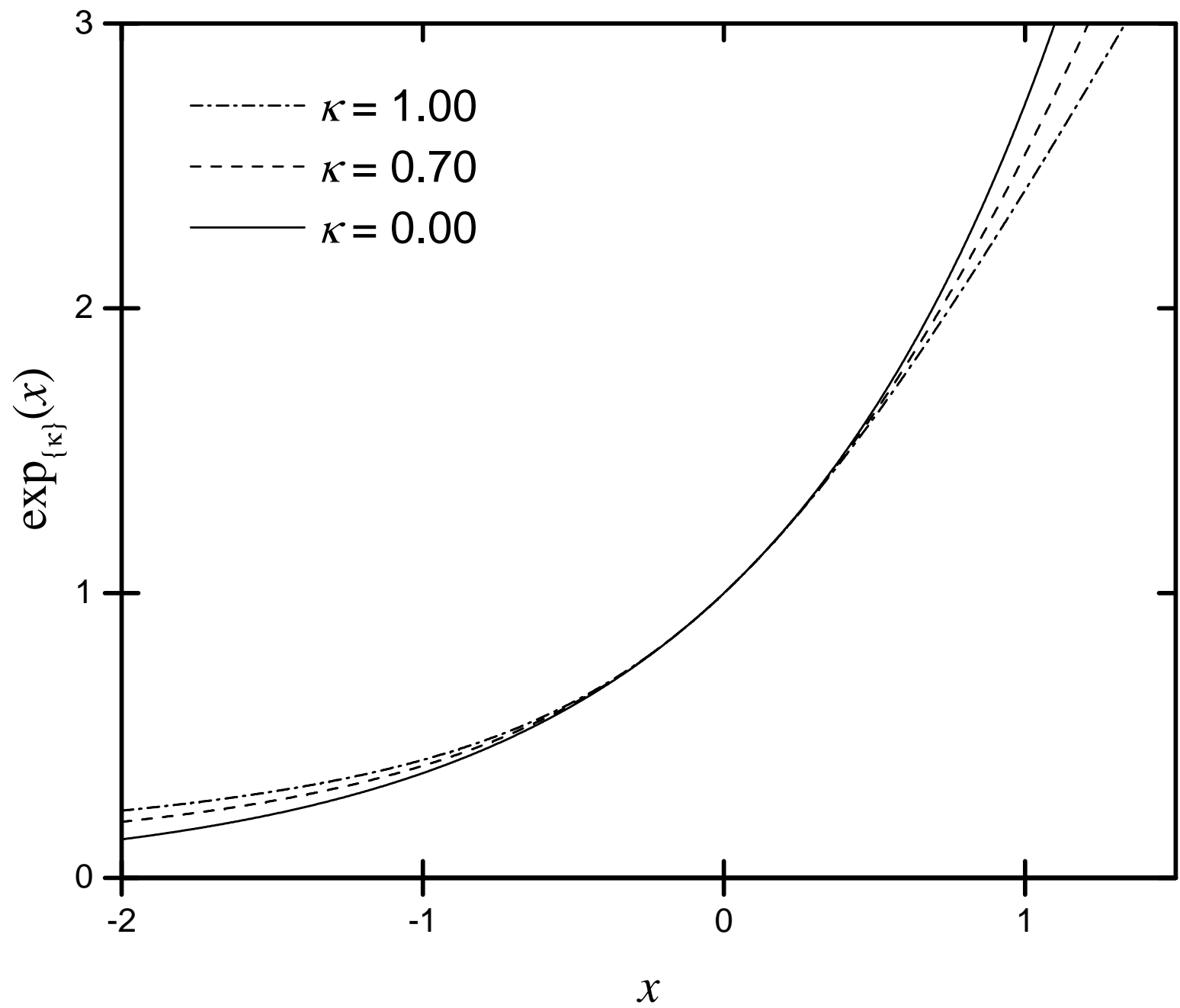
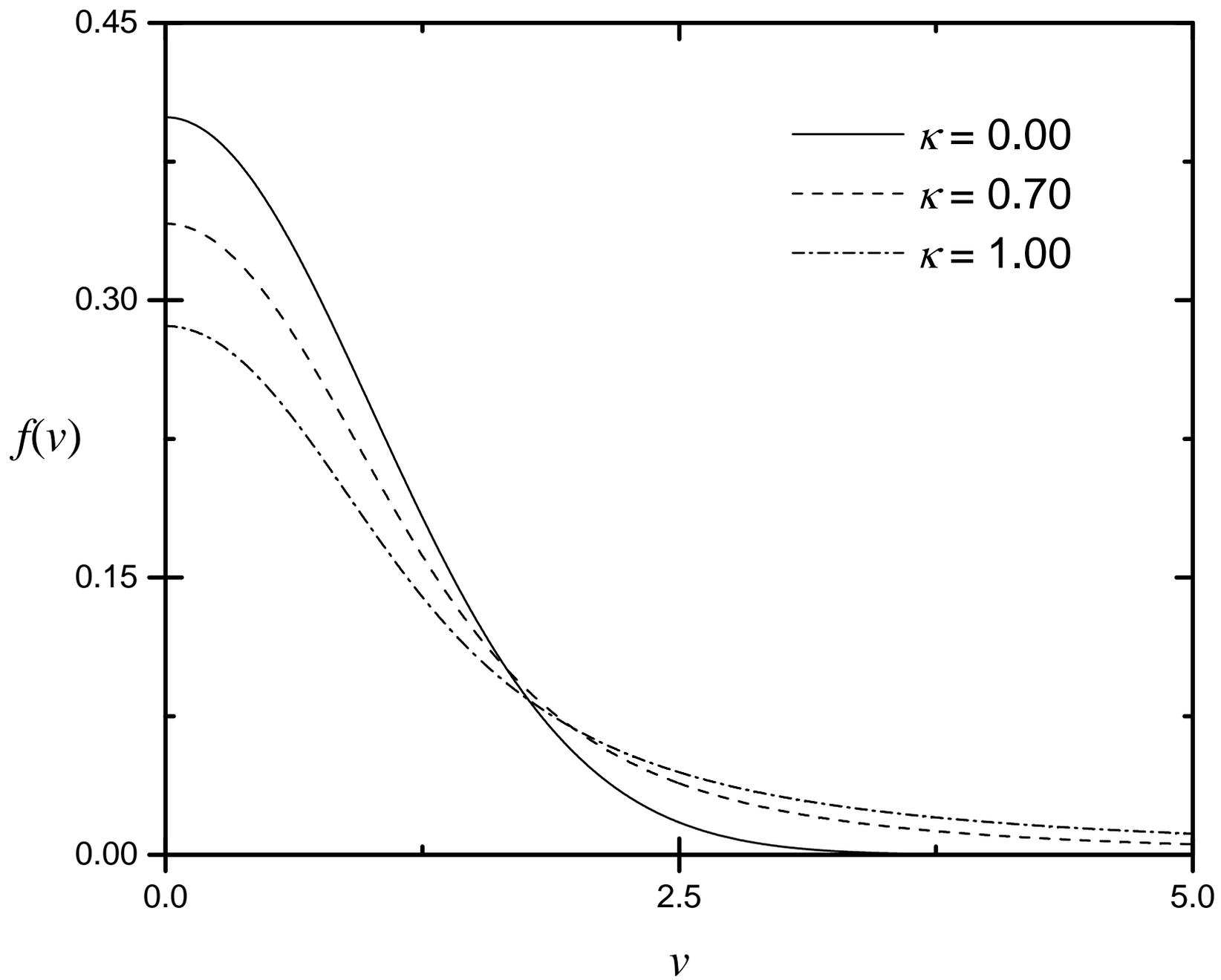
\end{document}